# Giant biquadratic interaction induced magnetic anisotropy in the iron-based superconductor $A_xFe_{2-y}Se_2$

Hai-Feng Zhu[1][2], Hai-Yuan Cao[1][2], Yun Xie[1][2], Yu-Sheng Hou[1][2], Shiyou Chen[3],

Hongjun Xiang[1][2], Xin-Gao Gong[1][2]

[1]Key Laboratory of Computational Physical Sciences (Ministry of Education), State Key

Laboratory of Surface Physics, Fudan University, Shanghai 200433, P. R. China

[2]Collaborative Innovation Center of Advanced Microstructures, Nanjing 210093, P. R. China

[3]Key Laboratory of Polar Materials and Devices (MOE), East China Normal University,

Shanghai 200241, China

The emergence of the electron-pocket only iron-based superconductor $A_xFe_{2-y}Se_2$ (A = alkali metal) challenges the Fermi-surface nesting picture established in iron-pnictides. It was widely believed that magnetism is correlated with the superconductivity in $A_xFe_{2-y}Se_2$. Unfortunately, the highly anisotropic exchange parameters and the disagreement between theoretical calculations and experimental results triggered a fierce debate on the nature of magnetism in $A_xFe_{2-y}Se_2$. Here we find that the strong magnetic anisotropy is from the anisotropic biquadratic interaction. In order to accurately obtain the magnetic interaction parameters, we propose a universal method, which does not need including other high energy configurations as did in conventional energy mapping method. We show that our model successfully captures the magnetic interactions in $A_xFe_{2-y}Se_2$ and correctly predicts the spin wave spectrum, in quantitative agreement with the experimental observation. These results suggest that the local moment picture, including the biquadratic term, can describe accurately the magnetic properties and spin excitations in $A_xFe_{2-y}Se_2$, which sheds new light on the future study of the high-$T_c$ iron-based superconductors.

## I. Introduction

The emergence of iron-selenide high-temperature (high-$T_c$) superconductors including $A_xFe_{2-y}Se_2$ (A = alkali metal)[1-6] and monolayer FeSe grown on $SrTiO_3$ substrates has attracted great interest[7-15]. They are the only known iron-based high-$T_c$ superconductors without a hole Fermi-surface[1]. In intensively studied iron-pnictides, the heavy electron-doping would suppress the hole Fermi-surface and keep the electron Fermi-surface, but they become non-superconducting because of diminished spin fluctuations [16]. However, in a phase-separated $A_xFe_{2-y}Se_2$ sample, the heavily electron-doped domain contributes to the superconductivity[1]. This alkali metal iron selenide with unique electronic structure and rather high-$T_c$ poses challenges to the physical picture previously established for iron-pnictides with both electron and hole Fermi-surface[5,6].

The spin-fluctuation, which is believed to be important to the superconducting pairing in iron-pnictides, comes from the Fermi-surface nesting between electron and hole pockets[17]. For this reason, the antiferromagnetism in iron-pnictides is described by the weak coupling model of the itinerant electrons[16]. In $A_xFe_{2-y}Se_2$, it has been suggested that magnetic ordering may arise from exchange interactions between localized moments, and the spin excitations can be well described by the $J_{1a}$-$J_{1b}$-$J_2$ Heisenberg model[18]. However, large anisotropic nearest-neighbor exchange constants have been observed in $AFe_{1.5}Se_2$, which cannot be described by the conventional Heisenberg model[18,19]. The origin of this highly anisotropic magnetic coupling is still a mystery. Nematic ordering, orbital ordering and biquadratic interactions were all proposed to be related to the high anisotropy of the magnetic coupling[20-25], but available experiments including the neutron scattering cannot determine which one plays the dominant role[18].

Previous first-principles calculations provided the exchange parameters[26,27], but the predicted spin excitations were substantially different from the fitting results in the inelastic neutron-scattering measurements[18]. For $KFe_{1.5}Se_2$, the magnetic interaction extracted from the conventional energy mapping method shows that the nearest-neighbor (NN) exchange parameters $J_{1a}$ and $J_{1b}$ are both antiferromagnetic, in contrast with the highly anisotropic sign-changing NN exchange couplings along the a and b axes ($J_{1a} > 0$, $J_{1b} < 0$) revealed by recent neutron-scattering experiments[18,26]. The energy mapping method usually works well in many systems with highly localized magnetic moments. However, the failure of the energy mapping method in $A_xFe_{2-y}Se_2$ questions whether the local moment picture is appropriate here, or if the conventional energy mapping method has some drawback when describing the magnetic interactions in $A_xFe_{2-y}Se_2$[28,29]. Resolving this problem would help us understand magnetism in high temperature iron-based superconductors. Since magnetism is considered to play a substantial role in iron-based superconductors, resolving the incomplete understanding of the magnetic properties in $A_xFe_{2-y}Se_2$ is crucial for further study of high-$T_c$ superconductivity in iron-based superconductors. Thus, it is natural to develop an appropriate model to describe the magnetism in $A_xFe_{2-y}Se_2$, which would (1) correctly predict the magnetic excitations consistent with the results from the neutron-scattering experiments, and (2) elucidate the origin of the highly anisotropic magnetic coupling.

Here, we formulate a new model, including the anisotropic biquadratic term which is found to be crucial. We show that by including biquadratic terms between the nearest neighbors, the experimentally observed magnetic properties can be accurately reproduced in different phases of $A_xFe_{2-y}Se_2$, without the next-next-nearest-neighbor exchange constant $J_3$. To obtain reliable effective exchange parameters, instead of the conventional energy mapping method, we propose a new convenient method based on non-collinear first-principle calculations. The new model gives the magnetic excitations, in the framework of density functional theory, which are in good agreement with the results observed by neutron scattering experiments. The present studies suggest that the highly anisotropic magnetic coupling originates from the non-negligible biquadratic term. Our results indicate that the localized model spin Hamiltonian with a biquadratic term well-describes the magnetic excitations in $A_xFe_{2-y}Se_2$. The success of the Heisenberg spin Hamiltonian including the biquadratic term reveals that magnetic interactions in $A_xFe_{2-y}Se_2$ are dominated by the localized spin moments. We have established a direct relationship between the microscopic electronic structure and the observed spin excitations from the neutron experiments, which sheds light on the future study of the high-$T_c$ iron-based superconductors considering the biquadratic magnetic interaction within the local moment pictures.

## II. Methods

### A. New method for computing the effective exchange parameters

The effective exchange parameters in the model Hamiltonian could be obtained through energy mapping. In principle, the fitted J should be independent on the details of the fitting process. However, for the conventional energy mapping method, a different choice of the high energy magnetic states in the mapping process would probably lead to different exchange parameters in some cases. This is due to the inclusion of meta-stable high energy states which might lead to different effective exchange parameters. With the four-state energy mapping method [30], the calculation of the exchange parameters in iron superconductors may also encounter convergence problems. The linear response method based on the perturbation theory obtained the effective exchange parameters in a different magnetic state of FeTe, but it fails to accurately predict the exchange parameters in $A_x Fe_{2-y} Se_2$.

To obtain reliable exchange parameters from our present method, we avoid using other high-energy magnetic states. Here we develop an efficient method based on non-collinear first-principles calculations. Our method computes the magnetic interaction only using states near the ground magnetic state, without including any other high energy states. It unambiguously addresses the effective exchange parameters and predicts the properties of the spin excitations accurately. It can deal with the contribution from both the Heisenberg term and the non-Heisenberg biquadratic term, based on efficient first-principles calculations.

Without loss of generality, we compute the magnetic exchange parameters based on a low-energy collinear spin state. Without loss of generality, we assume that the spins are along either the z or –z directions. Let us focus on a spin-pair between spin 1 and spin 2. The effective spin Hamiltonian is written as

$$E_{spin} = J_{12} \vec{S}_1 \cdot \vec{S}_2 - K \left( \vec{S}_1 \cdot \vec{S}_2 \right)^2 + \vec{S}_1 \cdot Q_1 + \vec{S}_2 \cdot Q_2 - S_1^2 \cdot Q_3 - S_2^2 \cdot Q_4 + E_{other} \qquad (1)$$

where $Q_1 = \sum_{i \neq 1,2} J_{1i} S_i$, $Q_2 = \sum_{i \neq 1,2} J_{2i} S_i$, $Q_3 = \sum_{i \neq 1,2} K_{1i} S_i^2$, $Q_4 = \sum_{i \neq 1,2} K_{2i} S_i^2$ and $E_{other}$ is the magnetic interaction between the spins without spin 1 and spin 2. It should be noted that $Q_1$, $Q_2$ and $E_{other}$ are independent of the directions of spin 1 and 2 since the magnetic order is a collinear spin state. This is also the case for $Q_3$ and $Q_4$. Then we considered four spin configurations of spin 1 and 2 (neglecting the spin-orbital coupling): (I) $S_1^x = 0$, $S_1^z = S$, $S_2^x = 0$, $S_2^z = S$; (II) $S_1^x = S \sin(\theta)$, $S_1^z = S \cos(\theta)$, $S_2^x = 0$, $S_2^z = S$; (III) $S_1^x = 0$, $S_1^z = S$, $S_2^x = -S \sin(\theta)$, $S_2^z = S \cos(\theta)$; (IV) $S_1^x = S \sin(\theta)$, $S_1^z = S \cos(\theta)$, $S_2^x = -S \sin(\theta)$, $S_2^z = S \cos(\theta)$. Here, we assume that spin 1 aligns ferromagnetically with spin 2 in the collinear spin state. In these four spin configurations, the direction of spins other than spin 1 and spin 2 always remain the same. If the initial state (I) is a collinear spin state like CAFM order, which is the common ground state for the iron-based superconductors, then the energies of the four states are

$$E_1 = J_{12}S^2 - K_{12}S^4 + Q_1S + Q_2S - Q_3S^2 - Q_4S^2 + E_{other} \qquad (2)$$

$$E_2 = J_{12}S^2\cos\theta - K_{12}S^4\cos(\theta)^2 + Q_1S\cos\theta + Q_2S - Q_3S^2\cos(\theta)^2 - Q_4S^2 + E_{other} \quad (3)$$

$$E_3 = J_{12}S^2\cos\theta - K_{12}S^4\cos(\theta)^2 + Q_1S + Q_2S\cos\theta - Q_3S^2 - Q_4S^2\cos(\theta)^2 + E_{other} \quad (4)$$

$$E_4 = J_{12}S^2\cos(2\theta) - K_{12}S^4\cos(2\theta)^2 + Q_1S\cos\theta + Q_2S\cos\theta \\ - Q_3S^2\cos(\theta)^2 - Q_4S^2\cos(\theta)^2 + E_{other} \qquad (5)$$

If the angle $\theta$ used in the calculation is small, then we can use the approximations $\cos\theta = 1 - \dfrac{\theta^2}{2}$ and $\sin\theta = \theta$. We then extract the contribution from the Heisenberg exchange coupling $J_{12}$ and the biquadratic term K.

$$J_{eff} = J_{12} - 2KS^2 = -\frac{E_1 + E_4 - E_2 - E_3}{\theta^2 S^2} \qquad (6)$$

If spin 1 is aligned antiferromagnetically with spin 2 in the collinear spin state, we obtain $J_{eff} = J_{12} + 2KS^2$.

The total energies of the different spin configurations are obtained directly from first-principles calculations, while the magnetic moment of the spin with a bias angle $\theta$ was stabilized with the help of an additional energy-penalty function. The method described above is a reasonable extension of the previous four-state energy-mapping analysis[30] that has been widely used to obtain the exchange coupling parameters. However, the essential difference is that the present method calculates the effective J for the specific spin configuration, without including the total energy of the high-energy spin configurations. The idea of our method is similar to that of the linear-response method, but in our method, the second derivatives of the total energy with respect to spin orientations are calculated in real space, which is much more convenient for the practical calculations. We have compared the magnetic exchange parameters J in $Cu_2OSeO_3$ obtained from our present method and the previous four-state energy mapping method, and the results from these two methods agree well with each other (see the supplementary material), which confirms that our new method describe the conventional Heisenberg model well.

To fit the $J_{1a,1b}$ and $K_{1a,1b}$ in $AFe_{1.5}Se_2$, we used the two magnetic states, the ground magnetic state A-collinear AFM order, and another metastable magnetic state P-collinear AFM order (see the Supplementary Materials). By fitting J and K in these two magnetic orders, we solved for $J_{1a,1b}$ and $K_{1a,1b}$ exactly from first-principles calculations.

## B. Computational details of DFT calculations

We employed the plane-wave basis and the projected augmented wave method[31] encoded in the Vienna *ab initio* simulation (VASP) package[32] to calculate the total energy as well as the electronic and magnetic properties. The generalized gradient approximation (GGA) with the Perdew-Burke-Ernzerhof (PBE) formula[33] was adopted for the exchange-correlation functional. We used a plane-wave cutoff energy of 500 eV and a Monkhorst-Pack mesh[34] of $6 \times 6 \times 6$ k points for structural optimization and $4 \times 4 \times 4$ k points for noncolllinear calculations with 0.1 meV Gaussian smearing in the calculation. A supercell of 24 Fe atoms, incorporating the experimental lattice parameters, was used to calculate the effective exchange parameters. For structural relaxation, all the inner atomic positions were fully optimized and the atoms were allowed to relax until the atomic forces were smaller than 0.01 eV/Å.

## III. RESULTS AND DISCUSSION

## A. Effective exchange parameters and the giant biquadratic interaction

First, we can accurately predict the experiment-fitted exchange parameters in $KFe_{1.5}Se_2$ with the collinear-AFM (CAFM) order with an effective model (Fig. 1). In order to describe the magnetic properties in $A_xFe_{2-y}Se_2$, we propose the following effective spin Hamiltonian with the biquadratic term:

$$H = J_{1a}\sum_{\langle ij \rangle_a} \vec{S}_i \cdot \vec{S}_j + J_{1b}\sum_{\langle ij \rangle_b} \vec{S}_i \cdot \vec{S}_j - \sum_{\langle ij \rangle_a} K_{1a}\left(\vec{S}_i \cdot \vec{S}_j\right)^2 - \sum_{\langle ij \rangle_b} K_{1b}\left(\vec{S}_i \cdot \vec{S}_j\right)^2 + J_2\sum_{\langle\langle ij \rangle\rangle} \vec{S}_i \cdot \vec{S}_j \ (7)$$

where $J_{1a,1b}$ are the in-plane nearest-neighbor (NN) Heisenberg exchange coupling parameters, $K_{1a,1b}$ are the anisotropic NN non-Heisenberg biquadratic coupling parameters, and $J_2$ is the next-nearest-neighbor (NNN) Heisenberg exchange coupling parameter. The summation in (7) are taken over the distinct pairs of lattice sites including the NN exchange couplings $J_{1a,1b}$, the next-nearest-neighbor (NNN) exchange coupling $J_2$, and the NN biquadratic couplings $K_{1a,1b}$, respectively. After optimizing the atomic positions, we use our new method to calculate the magnetic interactions in $KFe_{1.5}Se_2$ with the experimental lattice constant. The Heisenberg exchange parameters are $J_{1a}$ = 15.2 meV, $J_{1b}$ = 11.6 meV, $J_2$ = 14.7 meV, and the biquadratic interaction parameter $K_{1a}S^2$ = 5.8 meV, $K_{1b}S^2$ = 8.7 meV for $KFe_{1.5}Se_2$. We find the biquadratic term $K_{1a,1b}S^2$ is comparable to the Heisenberg term $J_{1a,1b}$, which is much more significant than that in other magnetic systems.

In order to compare with available results from experimental and theoretical studies, we use the relation $J_{1a'}$ = $J_{1a}$ + $2K_{1a}S^2$, $J_{1b'}$ = $J_{1b}$ − $2K_{1b}S^2$ [35] to calculate the effective $J_{1a'}$ and $J_{1b'}$, which is shown in Table I. The exchange parameters $J_{1b'}$ obtained from our new method is ferromagnetic, which is in good agreement with the experimental results[18], because the contribution from the biquadratic term $K$ is exactly included in the first-principles methods of our calculation. In previous $J_{1a}$-$J_{1b}$-$J_2$ models and energy mapping methods, the contribution from the non-Heisenberg term, like the biquadratic term $K$, had not been taken into consideration. For iron-based superconductors, the parent compounds are not Mott-insulators, so that the contribution from the itinerant electrons should also be considered in modeling the magnetic interaction[1]. From the above results, we see that the biquadratic term $K$ from the itinerant electrons in $KFe_{1.5}Se_2$ is too large to be neglected, and this giant biquadratic interaction could be the source of the strong anisotropic magnetic interaction for this system. Compared to the exchange parameters from previous energy mapping methods, our present method accurately describes the relative strength between the effective $J_{1a'}$, $J_{1b'}$ and $J_2$.

## B. Low energy magnetic excitation

The effective spin Hamiltonian and the obtained magnetic exchange parameters, allow us to give detailed descriptions of the spin wave and spin fluctuations. Due to the large effective magnetic moment in $KFe_{1.5}Se_2$ (around 2.8 $\mu B$ ), we assume S= 3/2 for each spin, and use the linearized Holstein-Primakoff (HP) transformation[27]. As shown in Fig. 2, the spin wave obtained from the present calculation is in excellent agreement with the results based on the experimentally fitted magnetic interactions. In the experiment, the less-dispersive gapped optical modes of the spin-wave are absent, because the signal of the optical modes is too weak to be distinguished from the background[18]. However, we provide a clear picture of the dispersion for both acoustic modes and optical modes. Since there are 6 Fe atoms in the magnetic unit cell, we can find 4 branches of optical modes and 2 branches of the gapless Goldstone modes in the spin wave spectrum. The spin anisotropy opens the gap in the Goldstone modes[27]. Both the optical modes and the Goldstone modes are doubly degenerate in the spin waves associated with the CAFM order. We also provide the spin wave computed from the experimental fitted exchange parameters. The dispersion of the spin wave obtained from our predicted exchange parameters combined with the HP transform shows excellent agreement with the spin wave compared to the experimental results.

Our model also reproduces the spin dynamic structure factor (SDSF). It is another important physical parameter measured in the neutron-scattering experiments[18], which contains information about spin correlations and their time evolution. Using the common leading-order approximation[27], we obtained the total dynamic spectral function in the entire Brillouin zone. The spin dynamic spectral function along the high symmetry lines is shown in Fig. 2. Note that the spin-wave spectrum is calculated in the magnetic unit cell while the SDSF is calculated in the atomic primitive unit cell. In inelastic neutron-scattering experiments, the Goldstone modes were observed around the $X$ point, $M$ point and along the $\Gamma - X'$ cut, while the intensity of the signal is strongest at the $X$ point. For this reason, only the SDSF around X point could be observed in the inelastic neutron-scattering experiment. The intensity of the signals from the optical modes is comparatively much weaker than that from the Goldstone modes, and the two branches of the optical modes were observed along the $X - M$ cut and the $M - X'$ cut, respectively. Due to the lack of experimental results for the optical-mode, our theoretical prediction of the optical-mode dispersion of $KFe_{1.5}Se_2$ could provide helpful hints for further study of the spin dynamics.

## C. Discussion

Finding the origin of the strong magnetic anisotropy is of great interest. The anisotropy of exchange parameters in $KFe_{1.5}Se_2$ is so strong that the signs of NN exchange parameters $J_{1a}$' and $J_{1b}$' are reversed. The sign-change NN exchange parameters are regarded as a common feature of the CAFM order in both iron-pnictides and iron-selenides[36]. In $KFe_{1.5}Se_2$, there are two origins for the strong magnetic anisotropy, including the rhombus ordered iron vacancy and the CAFM order. For further clarification of the contribution from the iron vacancy and the CAFM order to the biquadratic term, respectively, we calculate the exchange parameters in a metastable P-collinear AFM (PAFM) order with the same atomic structure of CAFM order[26,27]. The PAFM order is also a kind of collinear AFM order, but it exchanges the spin alignments along the $\hat{x}$ and $\hat{y}$ directions of the CAFM order. We find the $J_{1a'}$ and $J_{1b'}$ in the PAFM order are still anisotropic, but they are both antiferromagnetic couplings, which is different from those in the CAFM order. The anisotropy of the NN exchange parameters are also reduced in the PAFM, and the sign-changing phenomenon of $J_{1a'}$ and $J_{1b'}$ disappears (See the Supplementary Materials). The above results indicate that only the rhombus iron vacancy distribution itself could not be the cause of the strong magnetic anisotropy, and both the CAFM order and iron-vacancy in $KFe_{1.5}Se_2$ are necessary for the novel magnetic anisotropy observed in the $KFe_{1.5}Se_2$.

$AFe_{1.6}Se_2$ with $\sqrt{5} \times \sqrt{5}$ iron-vacancy is another major component in the phase-separated $A_xFe_{2-y}Se_2$ sample, which is also proposed to be the parent compound of the superconductivity. Several neutron-scattering experiments have been conducted to study the magnetic properties of $AFe_{1.6}Se_2$, but the unexpected block-AFM order and the complex magnetic interactions make it difficult and time-consuming to experimentally address the unambiguous magnetic excitation and exchange parameters of $AFe_{1.6}Se_2$[19]. On the other hand, it is also challenging for a theoretical approach to obtain the exchange parameters of $AFe_{1.6}Se_2$. The exchange parameters obtained from the conventional energy mapping method gives an unstable spin wave, which means that the conventional energy mapping method cannot give the proper magnetic interactions of $AFe_{1.6}Se_2$ as it fails in giving the right spin wave dispersion[37]. A linear-response method with the Green's function approach based on noncollinear density functional theory can predict a stable spin wave, but the details of the dispersion are qualitatively different from that of the neutron-scattering experiments[38]. Here we adopt our new method to calculate the magnetic interaction and the spin wave dispersion in $AFe_{1.6}Se_2$. The spin wave dispersion obtained by our method, as shown in Fig. 3, is not only qualitatively but also quantitatively in agreement with the experimental results. It should be noted that this is the first time to reproduce the experimental spin wave in this material through first principle calculations. The reproduction of the experimental spin wave indicates that our present method can be applied appropriately in the iron-chalcogenide superconductors, whose magnetic properties are well described by the local moment picture.

## IV. CONCLUSION

In conclusion, we have proposed a new method based on the non-collinear first-principles calculation to study the magnetic interaction and spin dynamics in iron-based superconductors $A_xFe_{2-y}Se_2$, where the prominent non-Heisenberg biquadratic interaction is exactly included in our model. Our method obtains the Heisenberg interaction and non-Heisenberg biquadratic interaction by performing convenient calculations on a magnetic order, without using any other meta-stable magnetic orders. We find that after taking the biquadratic term into account in the effective Hamiltonian, the experimentally observed strong anisotropy in $KFe_{1.5}Se_2$ is well described. By adopting our method, we can also accurately describe the spin wave and spin dynamic structure, which paves the way for the future experimental study of magnetic excitations in $A_xFe_{2-y}Se_2$ systems. Our present method based on the density functional theory, for the first time quantitatively captures the magnetic interaction and spin excitations in $A_xFe_{2-y}Se_2$ iron-based superconductors. It successfully builds the bridge between the theoretical Hamiltonian and the experimental observation of the magnetic interaction and the spin dynamics in iron-based superconductors, thus establishes unambiguously the local moment picture in $A_xFe_{2-y}Se_2$ systems. We believe that this method could be widely used in understanding magnetic interactions and spin dynamics of the iron-based superconductors like iron-selenide derived materials.

## Acknowledgements

We thank Professor J. Zhao for stimulating discussions. Work at Fudan was supported by NSFC, the Special Funds for Major State Basic Research, Research Program of Shanghai Municipality and MOE, Fok Ying Tung Education Foundation, and Program for Professor of Special Appointment (Eastern Scholar).

These authors contributed equally to this work.

Hai-Feng Zhu & Hai-Yuan Cao

**References:**

[1] E. Dagotto, *Rev. Mod. Phys.* **85**, 849 (2013).

[2] J. Guo, S. Jin, G. Wang, S. Wang, K. Zhu, T. Zhou, M. He, and X. Chen, *Phys. Rev. B* **82**, 180520(R) (2010).

[3] M.-H. Fang *et al.*, *Europhys. Lett.* **94**, 27009 (2011).

[4] T. Qian, X.-P. Wang, W.-C. Jin, P. Zhang, P. Richard, G. Xu, X. Dai, Z. Fang, J.-G. Guo, X.-L. Chen, and H. Ding, *Phys. Rev. Lett.* **106**, 187001 (2011).

[5] Y. Zhang *et al., Nat. Mater.* **10**, 273 (2011).

[6] D. Mou, S. Liu, X. Jia, J. He, Y. Peng, L. Zhao, L. Yu, G. Liu, S. He, X. Dong, J. Zhang, H. Wang, C. Dong, M. Fang, X. Wang, Q. Peng, Z. Wang, S. Zhang, F. Yang, Z. Xu, C. Chen, and X. J. Zhou, *Phys. Rev. Lett.* **106**, 107001 (2011).

[7] Q.-Y. Wang, Z. Li, W.-H. Zhang, Z.-C. Zhang, J.-S. Zhang, W. Li, H. Ding, Y.-B. Ou, P. Deng, K. Chang *et al.,Chin. Phys. Lett.* **29**, 037402 (2012).

[8] S. He, J. He, W. Zhang, L. Zhao, D. Liu, X. Liu, D. Mou, Y.-B. Ou, Q.-Y. Wang, Z. Li, L. Wang, Y. Peng, Y. Liu, C. Chen, L. Yu, G. Liu, X. Dong, J. Zhang, C. Chen, Z. Xu, X. Chen, X. Ma, Q. Xue, and X. J. Zhou, *Nat. Mater.* **12**, 605 (2013).

[9] S. Tan, Y. Zhang, M. Xia, Z. Ye, F. Chen, X. Xie, R. Peng, D. Xu, Q. Fan, H. Xu et al., *Nat. Mater.* **12**, 634 (2013).

[10] J. J. Lee, F. T. Schmitt, R. G. Moore, S. Johnston, Y.-T. Cui, W. Li, M. Yi, Z. K. Liu, M. Hashimoto, Y. Zhang, D. H. Lu, T. P. Devereaux, D.-H. Lee, and Z.-X. Shen, *Nature* **515**, 245 (2014).

[11] R. Peng, H. C. Xu, S. Y. Tan, H. Y. Cao, M. Xia, X. P. Shen, Z. C. Huang, C. H. P. Wen, Q. Song, T. Zhang, B. P. Xie, X. G. Gong, and D. L. Feng, *Nat. Commun.* **5**, 5044 (2014).

[12] R. Peng, X. P. Shen, X. Xie, H. C. Xu, S. Y. Tan, M. Xia, T. Zhang, H. Y. Cao, X. G. Gong, J. P. Hu, B. P. Xie, and D. L. Feng, *Phys. Rev. Lett.* **112**, 107001 (2014).

[13] J.-F. Ge, Z.-L. Liu, C. Liu, C.-L. Gao, D. Qian, Q.-K. Xue, Y. Liu, and J.-F. Jia, *Nat. Mater.* **14**, 285 (2015).

[14] H.-Y. Cao, S. Tan, H. Xiang, D. L. Feng, and X.-G. Gong, *Phys. Rev. B* **89**, 014501 (2014).

[15] H.-Y. Cao, S. Chen, H. Xiang, and X.-G. Gong, *Phys. Rev. B* **91**, 020504(R) (2015).

[16] D. C. Johnston, *Adv. Phys.* **59**, 803 (2010).

[17] P. Dai, J. Hu, and E. Dagotto, *Nat. Phys.* **8**, 709 (2012).

[18] J. Zhao, Y. Shen, R. J. Birgeneau, M. Gao, Z. Y. Lu, D. H. Lee, X. Z. Lu, H. J. Xiang, D. L. Abernathy, and Y. Zhao, *Phys. Rev. Lett.* **112**, 177002 (2014).

[19] M. Wang, C. Fang, D.-X. Yao, G. Tan, L. W. Harriger, Y. Song, T. Netherton, C. Zhang, M. Wang, M. B. Stone, W. Tian, J. Hu, and P. Dai, *Nat. Commun.* **2**, 580 (2011).

[20] A. L. Wysocki, K. D. Belashchenko, and V. P. Antropov, *Nat. Phys.* **7**, 485 (2011).

[21] P. Goswami, R. Yu, Q. Si, and E. Abrahams, *Phys. Rev. B* **84**, 155108 (2011).

[22] C. Fang, H. Yao, W. F. Tsai, J. P. Hu, and S. A. Kivelson, *Phys. Rev. B* **77**, 224509 (2008).

[23] C.-C. Chen, B. Moritz, J. van den Brink, T. P. Devereaux, and R. R. P. Singh, *Phys. Rev. B* **80**, 180418(R) (2009).

[24] W. Lv, J. Wu, and P. Phillips, *Phys. Rev. B* **80**, 224506 (2009).

[25] F. Krüger, S. Kumar, J. Zaanen, and J. van den Brink, *Phys. Rev. B* **79**, 054504 (2009).

[26] X. W. Yan, M. Gao, Z. Y. Lu, and T. Xiang, *Phys. Rev. Lett.* **106**, 087005 (2011).

[27] M. Gao, X.-W. Yan, and Z.-Y. Lu, *J. Phys. Condens. Matter* **25**, 036004 (2013).

[28] T. Yildirim, Physica C (Amsterdam) **469**, 425 (2009).

[29] J. K. Glasbrenner, I. I. Mazin, H. O. Jeschke, P. J. Hirschfeld, and R. Valentí, *arXiv*:1501.04946.

[30] Xiang, H *et al., Dalton Trans.* **42**, 823-853 (2013); Xiang, H *et al., Phys. Rev. B* **84**, 224429 (2011).

[31] P. E. Blochl, *Phys. Rev. B* **50**, 17953 (1994).

[32] G. Kresse and J. Furthmuller, *Phys. Rev. B* **54**, 11169 (1996).

[33] J. P. Perdew, K. Burke, and M. Ernzerhof, *Phys. Rev. Lett.* **77**, 3865 (1996).

[34] H. J. Monkhorst and J. D. Pack, *Phys. Rev. B* **13**, 5188 (1976).

[35] J. Hu, B. Xu, W. Liu, N.-N. Hao, and Y. Wang, *Phys Rev B* **85**, 144403 (2012).

[36] J. Hu and H. Ding, *Sci. Rep.* **2**, 381 (2012).

[37] C. Cao and J. Dai, *Phys. Rev. Lett.* **107**, 056401 (2011).

[38] L. Ke, M. van Schilfgaarde, and V. Antropov, *Phys. Rev. B* **86**, 020402(R) (2012).

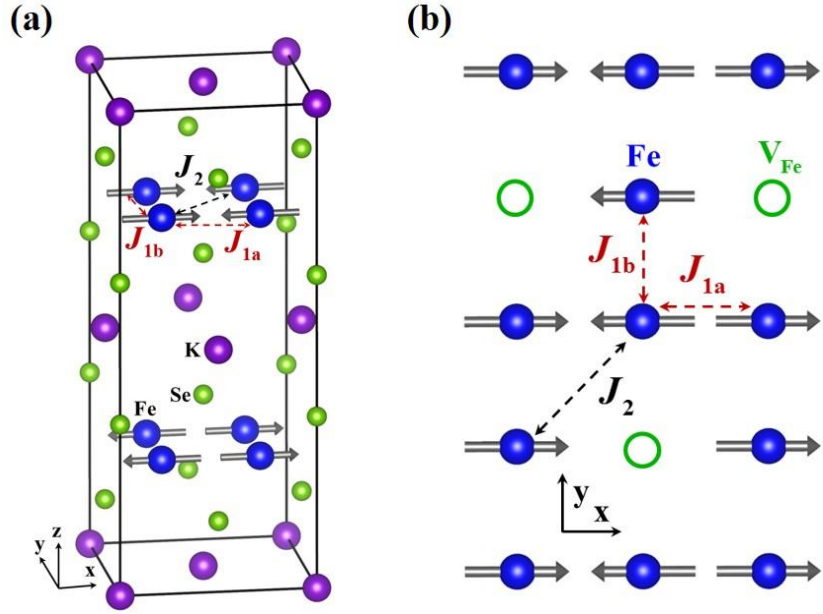

Figure 1: Magnetic structure for semiconducting $KFe_{1.5}Se_2$. Schematic diagram of (a) the stripe AFM order and (b) the order of iron vacancy in $KFe_{1.5}Se_2$. Here we assume the direction of antiparallel spin-alignment as $\hat{x}$ and that of parallel spin-alignment as $\hat{y}$.

-

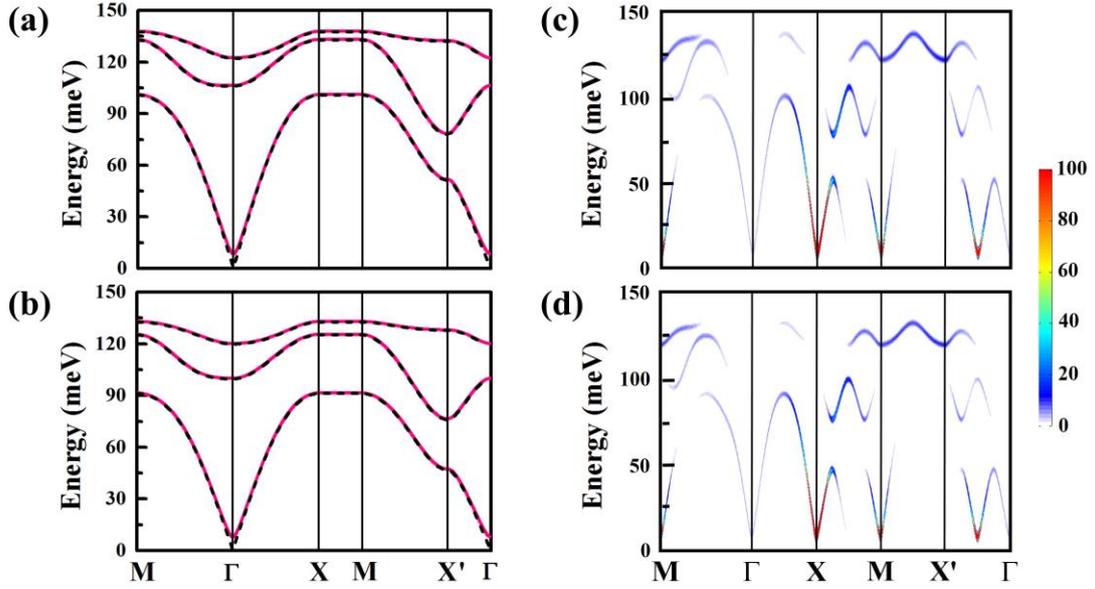

Figure 2: The spin-wave spectrum calculated by the exchange parameters obtained by (a) our new method and (b) the experimental fitting of KFe$_{1.5}$Se$_2$. The black dashed line (red solid line) denotes the spin-wave without (with) spin anisotropy. The spin dynamical structure factor (SDSF) of KFe$_{1.5}$Se$_2$ calculated with the exchange parameters obtained by (c) our present method and (d) the experimentally fitted parameters. The intensity of the SDSF has been rescaled to see the optical modes more clearly. Note that the spin-wave spectrum is calculated in the magnetic unit cell while the SDSF is calculated in the atomic unit cell.

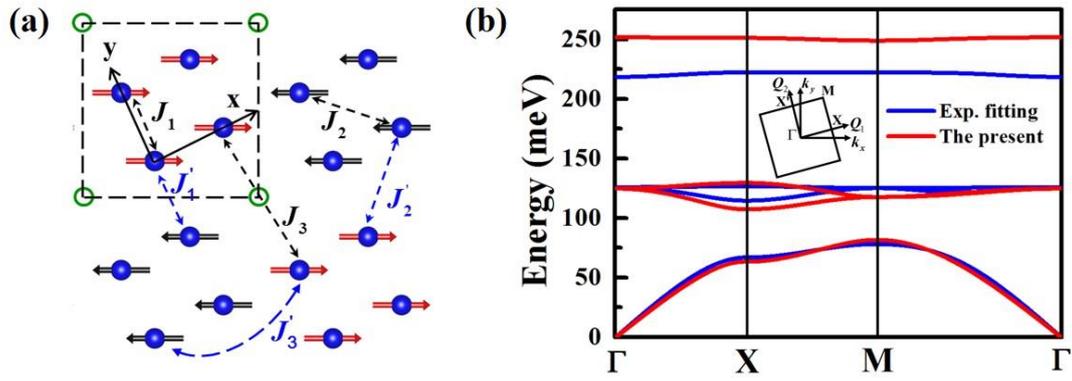

Figure 3: (a) Schematic diagram of the block-antiferromagnetic order of $KFe_{1.6}Se_2$. The square enclosed by the dashed lines denotes the atomic unit cell. (b) The spin-wave spectrum calculated with the exchange parameters from the present method (red line) and the experimental fit results (blue line). Here we assume S=2. Inset: the Brillouin zone of the magnetic unit cell in $KFe_{1.6}Se_2$.

| Method | $J_{1a}'$ (meV) | $J_{1b}'$ (meV) | $J_2$ (meV) |
|---|---|---|---|
| Present work | 26.7 | -5.8 | 14.7 |
| Expt. [Ref. 18] | 25.3 | -7.5 | 12.7 |
| Previous Cal. [Ref. 27] | 23.4 | 8.5 | 23.8 |

Table I: The effective exchange parameters $J_{1a}'$, $J_{1b}'$ and $J_2$ for KFe$_{1.5}$Se$_2$ with lattice constants from experimental data. The exchange parameters are obtained from our present method, the experimental fit to the spin wave spectrum and first-principles energy mapping. Here we assume S = 3/2.